# Spin momentum transfer in current perpendicular to the plane spin valves

M. Covington[a], A. Rebei, G. J. Parker, and M. A. Seigler

*Seagate Research, 1251 Waterfront Place, Pittsburgh, PA 15222*

We present experimental and numerical micromagnetic data on the effect of spin momentum transfer in current perpendicular to the plane spin valves. Starting from a configuration with orthogonal free and pinned layer magnetizations, the free layer magnetization exhibits abrupt current induced switching that is qualitatively consistent with the spin torque model. When operating the spin valve as a field sensor, spin transfer can produce a change in resistance that mimics an effective magnetic field and induce magnetic instability that requires a larger bias field in order to stabilize the device.

PACS #'s: 72.25.-b, 85.75.-d, 85.70.Kh

Magnetoresistive devices play a vital role as read sensors in hard disk drive recording heads. State-of-the-art readers are giant magnetoresistance (GMR) devices in which current flows in the plane (CIP) of the thin film multilayer. However, an alternate geometry, where current flows perpendicular to the plane (CPP), is being considered for higher areal density recording systems [1, 2]. There are many similarities between CIP and CPP readers, as both types rely on GMR to transduce magnetic flux to voltage, but one key difference is the occurrence of spin momentum transfer in CPP devices. This effect manifests itself at the nanometer scale dimensions relevant for future data storage applications, and it will be important to understand its full impact before implementing CPP reader technology.

Spin transfer refers to the exchange of spin angular momentum between conduction electrons and the magnetic moment of a ferromagnet [3, 4]. A spin polarized current flowing into a ferromagnet will produce a net torque that can induce either magnetization reconfiguration [5-9] or precession [10-15] for sufficiently large current densities. This has attracted interest in the potential application of spin transfer as a write mechanism in magnetic random access memory (MRAM) that can switch magnetic bits between bistable magnetic configurations. However, the design issues for readers in recording heads are different. The magnetization in a reader needs to passively sense the stray fields from the media rather than be actively driven by current, and it will be important to ensure a stable magnetic bias even in the presence of spin transfer torques. In this Letter, we address this issue by studying the impact of spin transfer on the magnetic stability of CPP spin valves as a function of current.

The CPP spin valves have a layer structure from bottom to top of IrMn70/CoFe60/Ru10/CoFe40/Cu22/CoFe30, where the numbers indicate the layer thicknesses in Å. The bottom CoFe layer is called the pinned layer (PL), the middle CoFe layer is the reference layer (RL), and the top CoFe layer is the free layer (FL). The PL and RL are strongly coupled antiferromagnetically across the Ru layer and are collectively referred to as a synthetic antiferromagnet (SAF). The CPP pillars have a circular cross-section with a diameter of ~100 nm and were etched such that approximately 20 Å of the PL material remained. The PL and antiferromagnet, IrMn, therefore have large volumes that are magnetically stable in the presence of Joule heating by the bias current. Finally, the device resistance, $R \equiv V/I$, is measured with the four-probe geometry shown in Figure 1a, where positive current flows from the bottom lead to the top lead.

Examples of the device characteristics for small bias currents are shown in Figure 1. The resistance versus magnetic field applied along the easy axis is shown in Figure 1b. Minor loops confirm that the switching at 80 Oe for the forward field sweep and -290 Oe for the negative sweep correspond to magnetization reversal of the FL. The switching at the other fields corresponds to the magnetization reversal of the entire pinned SAF structure. The large volume of the PL and the antiferromagnetic coupling to the RL results in the high resistance antiparallel state between the FL and RL at 1 kOe, although the resistance eventually decreases with increasing easy axis field as the RL magnetization rotates towards the field direction. The resistance versus hard axis field data in Figure 1c indicate that, aside from the hysteretic behavior at small fields, the FL is aligned along the field direction and the RL is roughly orthogonal to the FL for the field range shown.

When spin valves are used as readers, a magnetic field aligns the FL magnetization into an orthogonal state with respect to the RL magnetization. The first measurements pertain to the stability of this magnetic configuration and show that the configuration is stable for small currents but unstable above a field dependent current threshold. The FL magnetization was prepared by first applying a +3 kOe saturation field along the hard axis and then reducing the field to the value used when sweeping the current, which configured the device into an intermediate resistance state. Figure 2 shows examples of both hysteretic and non-hysteretic resistance versus current characteristics observed with these devices. We attribute the gradual increase in resistance with increasing current to Joule heating and the sharp jumps in resistance to a change in the FL configuration. The current densities that reconfigure the FL are on the order of $1 \times 10^8$ A/cm$^2$, and the difference between high and low resistance states, $\Delta R_I$, in the resistance versus current curves is slightly less than the magnetoresistance, $\Delta R_H$, from the resistance versus field curves. For Figures 2a and 2b, the $\Delta R_I$ values are 79% and 74%, respectively, of the $\Delta R_H$ values.

The resistance versus current data are consistent with spin transfer in the sense that electron flow from the FL to the RL favors antiparallel alignment and the opposite flow favors parallel alignment. Furthermore, the resistance versus field data preclude the possibility that the data in Figure 2 are due to a combination of nontrivial magnetic anisotropy and field induced switching. Measurements of the FL minor loop versus field angle indicate that there is a net unidirectional anisotropy aligned antiparallel to the RL magnetization direction, which is likely due to magnetostatic fields. A hard axis bias field simply adds an orthogonal unidirectional anisotropy. Field sweeps along the easy axis show that the FL minor loop crosses over from hysteretic 180° switching to smoothly varying, non-hysteretic rotation with increasing dc hard axis field bias. Finally, the data in Figure 2 are independent of which set of leads are used to pass current through the device. The in-plane fields generated by current flowing in the top and bottom leads are therefore negligible and can be excluded as a possible mechanism for the current induced reconfiguration.

We have also performed numerical micromagnetic calculations of the Landau-Lifshitz-Gilbert (LLG) equation with the spin torque term included, and the results further corroborate the interpretation of the data in Figure 2 in terms of spin transfer. The model is of a single 3 nm thick, 100x100 nm$^2$ ferromagnetic layer with a 10 Oe uniaxial anisotropy along the *x*-axis. A uniform hard axis bias field is along the *y*-axis. A uniform current density flows perpendicular to the layer along the *z*-axis, where the current is polarized along the *x*-axis. Analytic self-fields from this current are included assuming the leads are infinitely long with the same cross section as the ferromagnetic layer. The spin torque term is expressed in the same manner as in reference [16] with $a_J$=1 kOe for $j$=1x10$^8$ A/cm$^2$. The data in Figure 3 are for an effective saturation magnetization of $4\pi M_s$=12 kG, although similar results are obtained using the bulk CoFe $4\pi M_s$ of 18 kG. The change in resistance from the low resistance parallel state is assumed to obey the form $\Delta R = [1 - \cos\theta]/2$, where $\theta$ is the angle between the magnetization and the *x*-axis and $\cos\theta$ is averaged over the layer.

The numerical results for different LLG damping parameters, $\alpha$, and hard axis field values are shown in Figure 3. These data capture many of the same features observed experimentally. The numerical data display the same stable orthogonal magnetic configuration for small currents and abrupt current induced switching to parallel and antiparallel states. The theoretical curves also exhibit $\Delta R_I < \Delta R_H$ that is comparable to what is observed experimentally. The model indicates that this originates from the magnetization in the high and low resistance states being static and predominantly uniform, but canted away from the *x*-axis.

The qualitative agreement between experiment and theory is good, but there are quantitative differences in the region of the current induced realignment. The theoretical threshold currents can be adjusted by varying material parameters, such as $a_J$ and $\alpha$. But, an improved physical model is needed to better match the experimental field dependence and determine which factors, besides $\alpha$, produce hysteresis in the current induced switching. In terms of the dynamical response, the micromagnetic model predicts that the magnetization is stable above and below the threshold current, but exhibits persistent fluctuations in the region of the current induced reconfiguration. The spectral response of these dynamics covers a broad range of frequencies, and the error bars in Figure 3 show the range of resistance values that are representative for all of the numerical data in the transition region. The noise from these spin valves have been measured experimentally and 1/*f* noise, where *f* is the frequency, has been observed at frequencies of approximately 100 MHz in the vicinity of the threshold current. Ultimately, however, these issues are a subject for further study that is beyond the scope of this Letter.

Additional experiments have focused on the impact of spin transfer on read sensor operation, where there is a constant hard axis field and the easy axis field is varied. The two main effects that are observed are shown in Figure 4. The first is a change in resistance that grows with increasing current amplitude and shifts in a direction that depends on current polarity [17]. This change in resistance reflects a shift towards the magnetic configuration favored by the spin torque and is similar to the data in reference [6]. While the effect of spin transfer at high fields can be nontrivial, the FL minor loops that are relevant for readers indicate that a small shift in resistance occurs with only a minimal change in the slope of the resistance versus field, which mimics an effective easy axis magnetic field and can be easily accommodated within a recording system. However, the second effect illustrated by the data in Figure 4b reveals that spin transfer can destabilize a spin valve that is well behaved at low currents. The overall increase in resistance between the FL minor loops taken at +2 and -18 mA is due to Joule heating and is unimportant for magnetic recording. The important feature is that, for a given range of easy axis field variation, the minimum hard axis bias necessary to ensure magnetic stability depends on current amplitude. While larger hard axis fields can stabilize the device, this reduces the sensitivity of the device to easy axis fields.

In conclusion, we present experimental and numerical micromagnetic data that illustrate the effect of spin momentum transfer in CPP spin valves and, in particular, its impact on CPP read sensors. Spin transfer can overcome the unidirectional anisotropy produced by a hard axis bias field and realign the free layer magnetization. This spin transfer induced destabilization tends to be abrupt and stands in marked contrast to field induced reconfiguration. However, these current induced instabilities can be overcome by strengthening the magnetic biasing at the cost of decreasing the field sensitivity of the device.

We want to acknowledge the technical contributions of A. R. Eckert and the support of R. E. Rottmayer.

**Figure captions**

*Figure 1*. (a) Schematic cross-section of the spin valve. (b) and (c) Resistance versus field curves for a CPP spin valve acquired with a 1 mA bias current. Solid (dashed) lines indicate increasing (decreasing) field. (a) Field along the easy axis. (b) Field along the hard axis.

*Figure 2*. Examples of the two types of resistance versus current characteristics observed from CPP spin valves. The arrows indicate the direction the current is being swept. (a) Hysteretic realignment from the device in Figure1 biased at 0 Oe, as described in the text. (b) Non-hysteretic realignment from a different device. The hard axis field values are indicated by the legend.

*Figure 3*. Numerical micromagnetic data for a single ferromagnetic layer, as described in the text. The closed triangles and open circles represent the data for hard axis fields of 100 and 300 Oe, respectively. The error bars give an example of the minimum and maximum resistance values observed in the transition region. (a) Results for a damping coefficient of $\alpha=0.025$. (b) $\alpha=0.1$.

*Figure 4*. Resistance versus easy axis field, $H_{EA}$, with constant hard axis bias. Solid (dashed) lines represent increasing (decreasing) field. (a) Data for two opposite currents with a 625 Oe hard axis field. The heavy lines are overlapping minor loops taken from -500 to 500 Oe. (b) Minor loops acquired with a 375 Oe hard axis field for two different bias currents, as indicated.

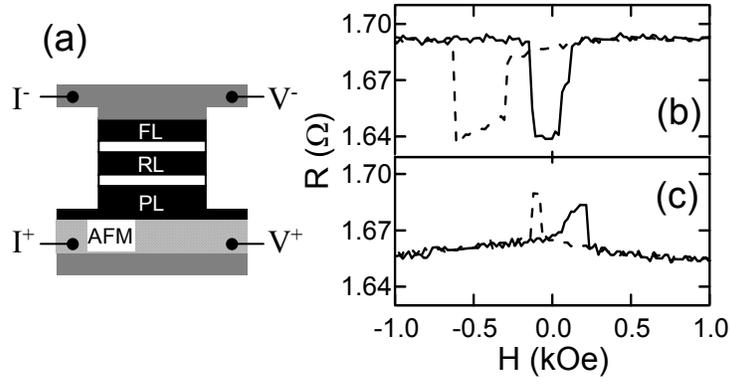

Figure 1

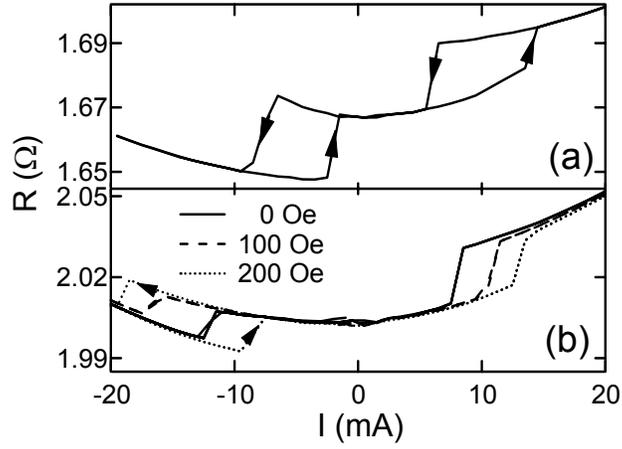

Figure 2

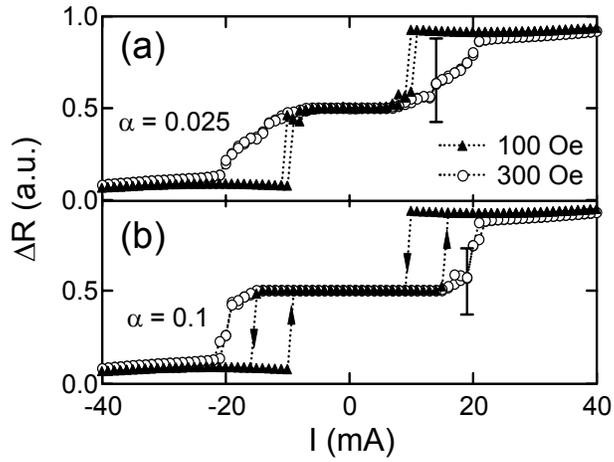

Figure 3

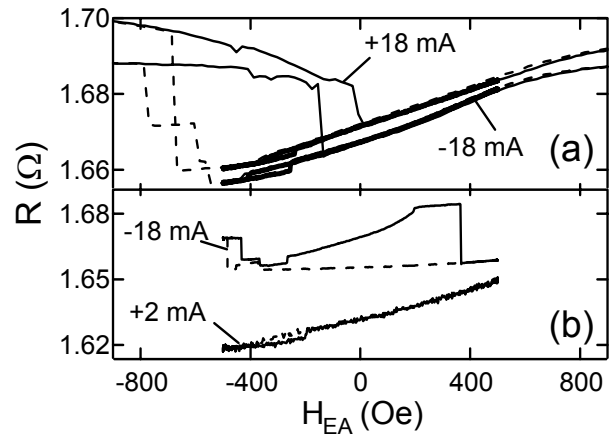

Figure 4